\documentclass[12pt]{article}

\usepackage{amssymb}
\usepackage{amsmath} 
\usepackage{latexsym}
\usepackage{amsthm} 
\usepackage{eucal} 

\newtheorem{thm}{Theorem}[section]

\newtheorem{prop}{Proposition} 
 
\newtheorem{rmk}{Remark}

\begin{document}

\title{Differential Geometry of Hydrodynamic Vlasov Equations}

\author{John Gibbons, Andrea Raimondo\\ \\ Imperial College\\ 180 Queen's Gate\\ London SW7 2AZ\\ j.gibbons@imperial.ac.uk, a.raimondo@imperial.ac.uk}



\date{}

\maketitle
\noindent

\begin{abstract}
We consider hydrodynamic chains in $(1+1)$ dimensions which are Hamiltonian
with respect to the Kupershmidt-Manin Poisson bracket. These
systems can be derived from single $(2+1)$ equations, here called {\em hydrodynamic
Vlasov equations}, under the map $A^n =\int_{-\infty}^\infty p^n f dp.$
For these equations an analogue of the Dubrovin-Novikov Hamiltonian structure
is constructed. The Vlasov formalism allows us to describe
objects like the Haantjes tensor for such a chain in a much more compact
and computable way. We prove that the necessary conditions found by Ferapontov
and Marshall in \cite{fema06} for the integrability of these hydrodynamic
chains are also sufficient. 
\end{abstract}

\section{Systems of hydrodynamic type}
\setcounter{equation}{0}

\emph{Systems of hydrodynamic type} are quasilinear first order PDE of the
form
\begin{equation}\label{fdhs}
u^i_t=v^i_j(u)u^j_x, \qquad i,j=1...N,
\end{equation}
where $(x,t)$ are the independent and $(u^1,...,u^N)$ the  dependent variables.
Here and below sums over repeated indices are assumed. 
A Hamiltonian formalism for systems of this type was introduced in \cite{duno83} by Dubrovin and Novikov, 
who defined a Poisson bracket of the form
\begin{equation}\label{dnpb}
\{I_\alpha,I_\beta\}=\int \frac{\delta I_\alpha}{\delta u^i(x)}\Pi^{ij}\frac{\delta I_\beta}{\delta u^j(x)}dx.
\end{equation}
\noindent
Here $I_\alpha,I_\beta$ are functionals of $u(x)$ 
and the first order differential operator $\Pi^{ij}$ is given by:
\begin{equation}\label{dnop}
\Pi^{ij}=g^{ij}(u)\frac{\partial}{\partial x}+b^{ij}_k(u) u^k_x.
\end{equation}
They showed that this is a Hamiltonian structure if 
$g^{ij}$ is a nonsingular contravariant metric and $b^{ij}_k=\!-g^{is}\Gamma^j_{sk}$,
where $\Gamma^i_{jk}$ is a symmetric connection of zero curvature that is compatible with the metric $g^{ij}$. It is immediate to see that a Hamiltonian of the form 
\begin{equation}\label{ham}
H=\int h(u)dx,
\end{equation}
where $h(u)$ is independent of $u_x,u_{xx},\dots$, together 
with the Hamiltonian structure (\ref{dnpb}),
leads to an equation of hydrodynamic type, specifically 
\begin{equation} \label{dneq}
u^i_t=\{u^i,H\}=\left(g^{ij}\partial_x+b^{ij}_k u^k_x\right)\frac{\partial
h}{\partial u^j}.
\end{equation} 
An obvious problem related with systems of hydrodynamic type (\ref{fdhs})
is to determine whether such a system is integrable, in the sense
that it admits infinitely many conserved densities and commuting flows;
in \cite{ts90},  Tsarev proved that this is true if the system is
hyperbolic and can
be written in diagonal form
\begin{equation*}
R^i_t=\lambda^i(R)R^i_x,
\end{equation*}
where $R^i$ are called the Riemann invariants and where the $\lambda^i$ (called
the characteristic velocities) satisfy the \emph{semi-Hamiltonian} condition
\begin{equation*}
\partial_k\left(\frac{\partial_j\lambda^i}{\lambda^j-\lambda^i}\right)=
\partial_j\left(\frac{\partial_k\lambda^i}{\lambda^k-\lambda^i}\right),
\end{equation*}
\noindent
where $\partial_k=\partial / \partial R^k.$
With these hypotheses, the system can then be integrated by the generalized hodograph
transformation (\cite{ts90}).  
We remark that the semi-Hamiltonian property is automatically satisfied for
a Hamiltonian system 
with Dubrovin-Novikov Hamiltonian structure, 
and that the conditions for the system to be respectively
diagonalizable, or semi-Hamiltonian, can be written invariantly; 
each corresponds to the vanishing of some tensor (\cite{pasvsh96},\cite{fema06}).  In particular, for the
diagonalizability condition, if one defines
the \emph{Nijenhuis tensor} of the matrix $v^i_j$ by:
\begin{equation*}
N^i_{jk}=v^s_j\frac{\partial v^i_k}{\partial u^s}-v^s_k\frac{\partial v^i_j}{\partial u^s}-v^i_s\left(\frac{\partial v^s_k}{\partial u^j}-\frac{\partial v^s_j}{\partial
u^k}\right),
\end{equation*}
and then the \emph{Haantjes tensor} by:
\begin{equation*}
H^i_{jk}=N^i_{\alpha\beta}v^\alpha_j v^\beta_k-N^\alpha_{j\beta}v^i_\alpha v^\beta_k- N^\alpha_{\beta k}v^i_\alpha v^\beta_j+N^\alpha_{jk}v^i_\beta
v^\beta_\alpha
\end{equation*}
then we have the following:
\begin{thm}\cite{ha55}
A hydrodynamic type system with mutually distinct characteristic speeds is
diagonalizable if and only if the corresponding Haantjes tensor vanishes identically. \end{thm}
\noindent

\subsection{Hydrodynamic chains}\label{hych}

\emph{Hydrodynamic chains} are defined as a natural generalization of
systems of hydrodynamic type, letting the number of variables and equations
go to infinity. More specifically, we consider, following Ferapontov and
Marshall (\cite{fema06}), systems of the type
\begin{equation}\label{hc}
A_t=V(A)A_x,
\end{equation}
where $A=(A^0,A^1,...)^t$ is an infinite column vector and $V(A)$
is an $\infty\times\infty$ matrix, with the following properties (see \cite{fema06},
\cite{fekhmapa06}),
\begin{itemize}
\item[1)] for every row only finitely many elements are nonzero
\item[2)] every element of the matrix depends only on a finite number of variables.
\end{itemize} 
\vskip 3mm
The variables $(A^0,A^1,...)$ are usually called \emph{moments}.  
The most famous example of a hydrodynamic chain is the Benney chain, 
\begin{equation}\label{benney}
A^n_t+A^{n+1}_x+n A^{n-1}A^0_x=0, \qquad n=0,1,\dots
\end{equation}
which was derived in \cite{be73} from the study of long nonlinear waves on
a shallow perfect fluid with
a free surface.  Kupershmidt and Manin (\cite{kuma77},\cite{kuma78})
found a Hamiltonian formulation,
\begin{equation*}
A_t^n=\left\{A^n,H\right\}_{KM}=\Pi^{nm}\frac{\delta H}{\delta A^m},
\end{equation*}
given by the Poisson operator 
\begin{equation}\label{kmpb}
\Pi^{nm}=(n+m)A^{n+m-1}\frac{d}{dx}+m A^{n+m-1}_x, \qquad n,m=0,1,\dots
\end{equation}
called the Kupershmidt-Manin bracket (KM bracket),
together with the Hamiltonian  
\begin{equation}\label{bham}
H=\int\left(\frac{1}{2}(A^0)^2+\frac{1}{2}A^2\right) dx.
\end{equation}
\vskip 7mm
\noindent
The KM bracket (\ref{kmpb}) has been considered (see, for example, \cite{do93}) as an infinite dimensional example of the Dubrovin-Novikov structure.
The general chain arising in this way, with a Hamiltonian density
\begin{equation}\label{hamden}
h(A^0,\dots, A^{N-1}),
\end{equation}
takes the form
\begin{equation}\label{ourch}
A^n_t=\sum_{m,l=0}^{N-1}(m+n)A^{m+n-1}h_{ml}A^l_x +\sum_{m=0}^{N-1}m h_m
A^{m+n-1}_x, \quad n=0,1,\dots
\end{equation}
\noindent
Here we have used the notation $h_i=\frac{\partial h}{\partial A^i}$, where $i=0,1,\dots.$
Compared with the finite dimensional case of Section 1, the theory of infinite
dimensional Poisson brackets of hydrodynamic type is not so well developed;
however, other examples of such Poisson brackets were given in \cite{ku83} as a generalization
of the Kupershmidt-Manin bracket (\ref{kmpb}), 
while the problem of the classification of such chains
has been approached recently in \cite{pa06}. The problem of finding integrable hydrodynamic chains was firstly approached in a systematic way
by Kupershmidt (\cite{ku83}), and, more recently and with different approaches, by Pavlov (\cite{pa06,pa206,pa306,pa406,pa506}),
Ferapontov and Marshall (\cite{fema06}), and Ferapontov, Khusnutdinova, Marshall and Pavlov (\cite{fekhmapa06}). 

Particularly, in
\cite{fema06}, the authors introduced an approach based on the Haantjes
tensor, generalizing Tsarev's results \cite{ts90}  for finite dimensional systems. For hydrodynamic chains, calculation of any one component
of this tensor only involves finite sums and hence is computable. Following this criterion, Ferapontov and Marshall considered Hamiltonian densities depending only on the first three moments
\begin{equation*}
h=h(A^0,A^1,A^2),
\end{equation*}
together with the KM bracket (\ref{kmpb}), and they
looked for the condition on the Hamiltonian for the system to have zero Haantjes
tensor. They found that the conditions 
\begin{equation*}
H^0_{jk}=0, \qquad j,k=0,1,2,\dots
\end{equation*}
on the first upper component give a complete system of ten third
order quasilinear partial differential equations, of which the simplest are
\begin{align}\label{pcond}
& h_{222}=\frac{5h_{22}^2}{2h_2}, \qquad h_{022}=\frac{5h_{02}h_{22}}{2h_2},
\qquad h_{122}=\frac{5h_{12}h_{22}}{2h_2}, \nonumber\\
& h_{002}=\frac{3h_{02}^2+2h_{00}h_{22}}{2h_2}, \nonumber\\
&\\
& h_{012}=\frac{3h_{02}h_{12}+2h_{01}h_{22}}{2h_2}, \nonumber\\
& h_{112}=\frac{3h_{12}^2+2h_{11}h_{22}}{2h_2}.\nonumber
\end{align}
The last four equations are much more cumbersome.  
\begin{rmk}
The list of equations above differs from the original paper \cite{fema06},
only in the names of the variables, as \cite{fema06} uses $u^n=A^{n-1}, n=1,2,\dots$
\end{rmk}
\vskip 5mm
\noindent
What Ferapontov and Marshall found were thus \emph{necessary} conditions
for the integrability of the chain. Remarkably, they were able to solve this
system; some of the the solutions they found correspond to known integrable
systems (correcting some errors in previous work), while the others, 
a much larger class, corresponded to systems since shown to be integrable. This result suggested a conjecture that the conditions above are not only necessary, but also sufficient.

\begin{rmk}
In a subsequent paper (\cite{fekhmapa06}), a similar problem was discussed,
but
with a Hamiltonian $H(A^0,A^1)$, and the $(\alpha\!-\!\beta)$ Hamiltonian structure,
\begin{equation}\label{alphabeta}
\Pi^{ij}_{(\alpha\beta)}=\big(\alpha(i+j)+\beta\big)A^{i+j-1}\frac{d}{dx}+\big(\alpha
j+\beta\big)A^{i+j-1}_x,
\end{equation}
which generalizes the KM bracket (\ref{kmpb}).
It is interesting to remark
that in this case too, the conditions for $H^0_{jk}=0$ give a complete set
of equations for the third derivatives of the Hamiltonian, and that for these systems it was shown that the conditions are indeed sufficient.
\end{rmk} 
\noindent
In the last section of this paper we will prove that the conditions
(\ref{pcond}) found by Ferapontov and Marshall are not only necessary for the vanishing
of the Haantjes tensor, but also sufficient. In order to do this, though, we need do
develop a somewhat different formalism.

\section{Vlasov formalism for Hamiltonian hydrodynamic chains}\label{sec3}

In this section we define the Vlasov equations, and we recall (\cite{gi81}) how a special
class of these equations can be related with hydrodynamic chains which
are Hamiltonian with respect to the Kupershmidt-Manin bracket. Moreover, we
show how to construct all the differential geometric objects related with
such chains in the Vlasov picture.
\vskip 5mm
\noindent
Let $f(x,p,t)$ be a distribution function in the in the $(1+1)$-dimensional phase space, and consider the Lie-Poisson bracket
\begin{equation}\label{lpb}
\{J,H\}_{LP}:=\iint f \left\{\frac{\delta J}{\delta f},\frac{\delta H}{\delta f}\right\}_{x,p}dp dx,
\end{equation}
where $H$, $K$ are functionals of $f$ and where the bracket $\{\,,\,\}_{x,p}$
is the canonical \lq single-particle' Poisson bracket. Hamilton's equations
related with such brackets,
\begin{equation*}
f_t=\{f,H\}_{LP},
\end{equation*}
or, equivalently,
\begin{equation*}
f_t+\left\{f,\frac{\delta H}{\delta f}\right\}_{x,p}=0,
\end{equation*} 
are called \emph{Vlasov equations}, and they arise in the theories of plasma
physics and vortex dynamics. The relation between these equations and the hydrodynamic chains
of the previous section is obtained by defining
\begin{align}\label{mom}
\mu\, :\, & f(p,x,t)\longmapsto \left\{ A^n(x,t)\right\}_{n=0}^\infty \\
& A^n=\int p^n f dp, \nonumber
\end{align}
where the integral above converges, for example, if $f$ is bounded and $|f| \rightarrow 0$ faster than $|p|^{-n}$, $\forall n\geqslant 1$.
\vskip 5mm
\noindent
As was shown in \cite{gi81} by one of the present authors, if we restrict
the bracket (\ref{lpb}) to functionals depending on the moments alone:
\begin{equation*}
H=H(A^0,\dots,A^{N-1})
\end{equation*}
then the Lie-Poisson bracket restricts to the Kupershmidt-Manin Poisson bracket (\ref{kmpb}):
\begin{equation*}
\{J,H\}_{LP}\equiv\{J,H\}_{KM}.
\end{equation*}
In order to prove this, is sufficient to use the chain rule for the map $\mu$:
\begin{equation}\label{chru}
\frac{\delta H(A^0,\dots A^{N-1})}{\delta f}=\sum_{n=0}^{N-1}\frac{\delta
H}{\delta
A^n} \frac{\delta A^n}{\delta f}=\sum_{n=0}^{N-1}\frac{\delta H}{\delta A^n}
p^n,
\end{equation}
then the KM bracket arises as the push forward of the L-P bracket
under this map. 
If we look at the evolution equations described by this bracket, for Hamiltonian functionals of type
\begin{equation}\label{hyfun}
H=\int h(A^0,\dots A^{N-1}) dx,
\end{equation}
we obtain a relation between a class of Vlasov equations 
\begin{equation}\label{hve}
f_t=\left\{f,H\right\}_{LP},
\end{equation}
and the Kupershmidt-Manin hydrodynamic chains (\ref{ourch})
\vskip 4mm
\begin{equation*}
A_t^n=\left\{A^n,H\right\}_{KM}, \quad n=0,1,\dots.
\end{equation*}
\vskip 5mm
\noindent
We call equations (\ref{hve}) \emph{hydrodynamic Vlasov equations}; more
explicitly, recalling that for functionals of type (\ref{hyfun}) we have
\begin{equation}
\frac{\delta H(A^0,\dots A^{N-1})}{\delta f}=\sum_{n=0}^{N-1}h_n p^n,
\end{equation}
these equations take the form
\begin{equation}\label{exhve}
f_t=\!\!\left(\sum_{n,m=0}^{N-1}\!p^n h_{nm} A^m_x \right)\!f_p
-\!\left(\sum_{n=0}^{N-1}n p^{n-1}h_n\!\right)\!f_x.
\end{equation}

\vskip 5mm 
\noindent
Ferapontov and Marshall started to study the
differential-geometric properties of such hydrodynamic chains using
the countably infinite set of discrete coordinates $A^n,$ but instead, it
is possible to study these properties by looking at the
corresponding hydrodynamic Vlasov equations directly. 
\vskip 5mm
\noindent
In order to do so, we want to consider equation (\ref{exhve}) as a kind of
$(1+1)$-dimensional hydrodynamic type system (\ref{fdhs}). Indeed, we notice that equations (\ref{exhve}) are linear with respect to the derivatives $f_t$ and $f_x$;
thus, we consider the function 
\begin{equation*}
f(p,x,t),
\end{equation*}
as a vector; the independent variable $p$ is treated as a continuous parameter, analogous
to the discrete index in the components of a finite dimensional vector. We will, for brevity,
suppress the dependence on $(x,t)$. In this way, the hydrodynamic Vlasov equations
(\ref{exhve}) can be viewed as $(1+1)$ hydrodynamic type systems of continuously
infinitely many equations and variables; indeed they can be written in the
form
\begin{equation}\label{chs}
f_t(p)=\int V\!\left({}^p_q\right)f_x(q) dq,
\end{equation}
where the kernel $V\!\left({}^p_q\right)$ is given by
\begin{equation}\label{ker}
V\!\!\left({}^p_q\right)\!\!=\!\!\left(\sum_{n,m=0}^{N-1}\!\!p^n q^m
h_{nm}\!\right)\!\!f_p-\!\left(\sum_{n=0}^{N-1}\!\!n
p^{n-1}h_n\!\right)\!\!\delta(p-q).
\end{equation}
\vskip 5mm
\noindent
It is important here that we do not consider (\ref{exhve}) as a $(2+1)$-dimensional
hydrodynamic type system. Rather, we consider $f_p$ as a functional of $f$, namely
\begin{equation*}
K[f]:=\int f(r)\delta'(p-r)dr=f_p,
\end{equation*}
where $\delta'(p-r)$ is the derivative of the Dirac delta function. Thus,
the kernel (\ref{ker}) may be considered as depending on $f$, analogously
to discrete non-linear hydrodynamic type systems.
\vskip 10mm
\noindent
To complete our construction, we need to substitute, in a formal
way, discrete objects with continuously indexed ones; namely:

\begin{center}
\begin{tabular}{c|c}
$n=0,1,2,\dots$ & $p\in \mathbb{R}$\\
\hline
&\\
$A^n(x,t)$ & $f(p,x,t)$ \\
&\\
$\frac{\partial h(A^0,\dots,A^{N-1})}{\partial A^n}$ &
$\frac{\delta h[f]}{\delta f(p)}$ \\
& \\
Sums on repeated & Integrals on repeated \\
discrete indices & continuous indices
\end{tabular}
\end{center}

\vskip 10 mm
\noindent
Using these coordinates, we can construct any tensor object related
with a hydrodynamic chain (\ref{ourch}), the relation being an
analogue of the classical change of coordinates of a tensor under
the map $\mu$.
The advantage of this formulation is that, instead of studying
infinite-component tensors, we can consider integral
operators, which are much more compact and computable. 
\vskip 5mm
\noindent
As an example, we write down explicitly the Vlasov formalism for the\\
Kupershmidt-Manin structure (\ref{kmpb}). This is given by the metric
\begin{equation*}
G^{mn}[A^0\!,\!A^1\!,\dots]=(m+n)A^{m+n-1},
\end{equation*}
\noindent
and to this metric corresponds, in the Vlasov coordinate, an operator $g[f]$, depending on two real parameters
\begin{equation}\label{metrics}
g^{(p,q)}[f], \qquad p,q\in \mathbb{R}
\end{equation}
and symmetric with respect to $p,q$. The relation with the metric in the Vlasov coordinates is the identity
\begin{equation}\label{change}
\iint \! g^{(p,q)}[f]\frac{\delta
A^m}{\delta f(p)}\frac{\delta A^n}{\delta f(q)} dp dq\!=\!G^{mn}[A^0\!,\!A^1\!,\dots]
\end{equation}
which is an analogue of the classical change of variables of a $(2,0)-$tensor under the map (\ref{mom}). Of course, when a continuous index is repeated,
we integrate with respect to the repeated index. 
If we take 
\vskip 3mm
\begin{equation}\label{met}
g^{(p,q)}[f]=-f_p(p)\delta(p-q),
\end{equation}
\vskip 3mm
\noindent
then, substituting in (\ref{change}), we indeed have
\vskip 2mm
\begin{equation*}
\iint g^{(p,q)}[f]p^m q^n dp dq=-\int f_p(p)p^{m+n}dp=(m+n)A^{m+n-1}.
\end{equation*}
\vskip 5mm
\noindent
We notice that, in the new coordinates, the Dirac delta function plays the
role of the Kronecker delta $\delta^i_j$, indeed, we have
\begin{equation*}
\sum_j\frac{\delta A^j}{\delta f(p)} \delta^i_j =\int \frac{\delta A^i}{\delta f(q)}\delta(p-q)dq.
\end{equation*}
As a consequence of this, we notice that the metric (\ref{met}) in the Vlasov coordinates has
diagonal form. In addition,
we will say that a metric $g^{(p,q)}$ is \emph{non degenerate} if
there exists an inverse metric $g_{(p,q)}$ such that
\begin{equation}\label{inv}
\int g_{(p,\alpha)}g^{(\alpha,q)}d\alpha=\delta(p-q).
\end{equation}
\vskip 5mm
\noindent
Continuing the analogy, we can now pursue a direct computation of the differential
geometric object we need, directly in the Vlasov coordinates. So, we define the \emph{Christoffel symbols}, which are given, in components, by the
following formula:
\begin{equation}\label{christ}
\!\!b{p\choose q,r}\!:=\!\frac{1}{2}\int
\!g^{(p,\alpha)}\left(\frac{\delta g_{(\alpha,q)}}{\delta f(r)}+
\frac{\delta g_{(\alpha,r)}}{\delta f(q)}-\frac{\delta
g_{(q,r)}}{\delta f(\alpha)}\right)d\alpha.
\end{equation}
For the metric (\ref{met}), the Christoffel symbols are
\begin{equation*}
\!\!b{p\choose q,r}\!=-\delta'(p-q)\delta(r-q),
\end{equation*}
and then the curvature, defined in analogy with the classical case as
\begin{equation}\label{curv}
R{s \choose p, q, r}\!:=\! \frac{\delta b{s \choose p, r}}{\delta f(q)}-\frac{\delta
b{s \choose p, q}}{\delta f(r)}\!+\!\int\! b{s \choose q, \alpha}b{\alpha \choose p, r}d\alpha\!-\!\int\! b{s \choose r, \alpha}b{\alpha \choose p, q}d\alpha 
\end{equation}
\vskip 5mm
\noindent
is found to be identically zero.

\vskip 50pt

\section{The Haantjes tensor for hydrodynamic Vlasov equations}

We introduce now the Nijenhuis and Haantjes tensors for a Vlasov
equation of hydrodynamic type (\ref{hve},\ref{exhve}). Particularly, in the
second part of the section we will consider the special case
when the Hamiltonian density depends only on the first three moments, $h(A^0,A^1,A^2),$
so that
\begin{equation*}
\frac{\delta h}{\delta f}=h_0+ p h_1+p^2 h_2,
\end{equation*}
and we calculate the conditions
for a system with such a Hamiltonian to have vanishing Haantjes tensor. As in the
previous examples, this differential geometric result for a Vlasov hydrodynamic
equation can be lifted to the corresponding hydrodynamic chain. Consider
first the general case of a Hamiltonian function of type (\ref{hamden}).
In order to simplify our notation, we write the kernel (\ref{ker}) as
\begin{equation}\label{kerker}
V{p \choose q}=B(p,q)f_p-A(p)\delta(p-q),
\end{equation}
where 
\begin{equation*}
A(p):=\sum_{n=0}^{N-1}nh_n p^{n-1}\quad, \quad B(p,q):=\sum_{n,m=0}^{N-1}h_{nm} p^n q^m 
\end{equation*}
are polynomials in $p$ and $p,q$ respectively, whose coefficients are the derivatives
of the Hamiltonian. As with the discrete case, we define the Nijenhuis tensor for
a hydrodynamic Vlasov equation as
\begin{equation}\label{vnij}
N{p\choose q,r}\!\!\!:=\!\!\!\int\! V\!{\alpha\choose q}\frac{\delta V{p\choose r}}{\delta
f(\alpha)}-V\!{\alpha\choose r}\frac{\delta V{p\choose q}}{\delta
f(\alpha)}- V\!{p\choose \alpha}\!\!\left(\frac{\delta V\!{\alpha\choose r}}{\delta f(q)}\!-\!\frac{\delta V\!{\alpha\choose q}}{\delta f(r)}\right)\!d\alpha,
\end{equation}
while the Haantjes tensor is then given by
\begin{align}\label{vhaa}
H{p\choose q,r}\!:=&\! \iint\!\!\left( N\!{p\choose \alpha,\beta}V\!{\beta\choose q}V\!{\alpha\choose r}\!-\!N\!{\alpha\choose\beta ,r}V\!{p\choose \alpha}V\!{\beta\choose q}+\right.\nonumber\\
&\left. -N\!{\beta\choose
q ,\alpha}V\!{p\choose \beta}V\!{\alpha\choose r}\!+\!N\!{\beta\choose q ,r}V\!{p\choose \alpha}V\!{\alpha\choose \beta}\right)d\alpha d\beta.
\end{align}

\vskip 5mm
\noindent
Let us calculate
the Nijenhuis tensor for a general kernel (\ref{kerker}); first of all, we
have to compute the variational derivative of $V$ with respect to $f$. A
direct calculation shows that 
\begin{align}
\frac{\delta V{p\choose q}}{\delta f(r)}=&\left(\sum_{n,m,l=0}^{N-1}p^n q^m
r^l h_{nml}\right)f_p+\left(\sum_{n,m=0}^{N-1}p^n q^m h_{nm}\right)\delta'(p-r)+\nonumber\\
&-\left(\sum_{n,m=0}^{N-1}np^{n-1}q^m h_{nm}\right)\delta(p-q).\label{delv}
\end{align}
\noindent
If we denote
\begin{equation*}
C(p,q,r):=\sum_{n,m,l=0}^{N-1}p^n q^m r^l h_{nml},
\end{equation*}
then the identity (\ref{delv}) may be written
\begin{equation}\label{difker}
\frac{\delta V{p\choose q}}{\delta f(r)}=C(p,q,r)f_p+B(p,q)\delta'(p-r)-\frac{\partial
B(p,q)}{\partial p}\delta(p-q).
\end{equation}
Substituting equations (\ref{kerker}) and (\ref{difker}) into the definition
of the Nijenhuis tensor (\ref{vnij}), and using properties of the delta function, we obtain
\begin{equation*}
N{p\choose q,r}=E(p,q,r)f_p+F(p,r)\delta(p-q)-F(p,q)\delta(p-r),
\end{equation*}
where $E$ and $F$ are polynomials given by
\begin{equation*}
\!F(p,q)\!=\!\left(\!A(q)\!-\!A(p)\right)\frac{\partial B(p,q)}{\partial p}\!+\!\frac{\partial
A(p)}{\partial p}B(p,q)\!-\!\!\int\!\! B(\alpha,q)\frac{\partial B(p,\alpha)}{\partial p}\!f_\alpha  d\alpha
\end{equation*}
and
\begin{align*} 
\!\!E(p,q,r)\!=\! &\left(\!A(r)\!-\!A(q)\right)\!C(p,q,r)\!+\!B(p,q)\!\frac{\partial B(p,r)}{\partial
p}\!-\!B(p,r)\!\frac{\partial B(p,q)}{\partial p}+\nonumber\\
&+\!B(q,r)\!\left(\frac{\partial B(p,q)}{\partial q}-\frac{\partial B(p,r)}{\partial r}\right).
\end{align*}

\vskip 7mm
\begin{rmk}\label{rmkint}
It is easy to verify that $E$ is a polynomial in $p,q,r$ whose coefficients
are quadratic expressions in the derivatives of $h$, for it is defined as a product of polynomials which are linear in the derivatives of $h$. For $F$, though, this fact is less clear, because of the integral in the last
term. However, it is possible
to write the integrand as a polynomial in $\alpha$, since we have
\begin{align*}
\int\! B(\alpha,q)\frac{\partial B(p,\alpha)}{\partial p}f_\alpha
d\alpha=&\!\int\!\sum_{n=0}^{2N-2} P_n(p,q)\alpha^n f_\alpha d\alpha=\\
=&-\!\sum_{n=0}^{2N-2}P_n(p,q)\!\!\int\! n \alpha^{n-1}f(\alpha)d\alpha=\\
=&-\sum_{n=0}^{2N-2}n P_n(p,q)A^{n-1}.
\end{align*}
Here the $P_n$ are suitable polynomials with coefficients quadratic in the derivatives of the Hamiltonian. We observe that the number of moments appearing
in these expressions will generally be bigger than $N$. Similar dependence
on the $A^n$ will appear in the calculation of the Haantjes tensor as well.
\end{rmk}

\vskip 7mm
\noindent
The calculation of the Haantjes tensor is similar. It follows from the above, with a long but essentially straightforward calculation, that
\begin{equation}\label{vlhaa}
H{p\choose q,r}=Q(p,q,r)f_p.
\end{equation}
\vskip 5mm
\noindent
We will call the polynomial $Q(p,q,r)$, above, the \emph{Haantjes polynomial} for
the related hydrodynamic Vlasov equation. Remarkably, in the above expression
there do not appear any coefficients in the $\delta$-function or its derivative.
It would be interesting to find a deeper explanation for this. 
In addition, given the Haantjes tensor $H^i_{jk}$ for the corresponding  hydrodynamic chain,
we have the following relation:
\begin{equation*}
\sum_j\sum_k H^i_{jk}\frac{\delta A^j}{\delta f(q)}\frac{\delta A^k}{\delta f(r)}=\int H{p\choose q,r}\frac{\delta A^i}{\delta f(p)} dp,
\end{equation*} 
which is the change of variables under the map (\ref{mom}), introduced in
Section (\ref{sec3}), for a tensor of type $(1,2)$. Explicitly:
\begin{equation}\label{hrel}
\sum_{j,k}q^jr^k H^i_{jk}=\int p^i Q(p,q,r)f_p dp.
\end{equation}
\vskip 5mm
\noindent
So, to study the properties of the Haantjes tensor of a chain it is sufficient to study
the properties of the corresponding Haantjes polynomial.
It is possible to show that, for
$N>2$, this polynomial has the form
\begin{equation}\label{qexpl}
Q(p,q,r)=\sum_{l=0}^{4(N-2)}\sum_{m=0}^{3N-5}\sum_{n=0}^{3N-5}Q_{lmn}p^l
q^m r^n,
\end{equation}
where the coefficients $Q_{lmn}$ are linear or quadratic expressions of
type
\begin{equation*}
Q_{lnm}=Q_{lnm}\left(h_i,h_{ij},h_{ijk},A^0,\dots,A^{4N-7}\right)\qquad i,j,k=0,\dots,N-1,
\end{equation*}
involving the first,
second and third derivatives of the Hamiltonian $$h(A^0,\dots A^{N-1}),$$ as
well as explicit dependence on the moments
$$A^0,\dots A^{N-1},$$
and on \lq extra' moments not appearing in the Hamiltonian, 
\begin{equation}\label{extramom}
A^N,\dots A^{4N-7}.
\end{equation}
\noindent
These appear, as explained in Remark \ref{rmkint}, when integrals of the
form $\int \alpha^nf_\alpha d\alpha$ are evaluated.  The Haantjes polynomial is antisymmetric with respect to $q$ and $r$.
Writing the Haantjes polynomial $Q$ as in (\ref{qexpl}), the equation (\ref{hrel}) becomes
\begin{align*}
\sum_{j,k=0}^\infty q^jr^k H^i_{jk}\!=&\!\!\int\! p^i \!\left(\sum_{j,k=0}^{3N-5}\sum_{l=0}^{4(N-2)}Q_{ljk}p^l
q^j r^k\right)\! f_p dp =\\
=&\sum_{j,k=0}^{3N-5}\sum_{l=0}^{4(N-2)}\left(\! Q_{ljk}\!\int p^{i+l}f_p dp\right)q^jr^k,
\end{align*}
so that
\begin{equation*}
H^i_{jk}=-\!\!\sum_{l=0}^{4(N-2)}(i+l) Q_{ljk}A^{i+l-1}.
\end{equation*}
\vskip 7mm
\noindent
As a consequence of the equations above, we have, for every fixed $i$, that \begin{equation*}
H^i_{jk}=0 \qquad \forall j,k>3N-5.
\end{equation*}

\noindent
This fact, noticed in \cite{fema06} by Ferapontov and Marhall, in
this setting turns out to be a straightforward consequence of the dependence of the
Hamiltonian on finitely many moment variables. In order for the Haantjes
tensor to vanish identically, we note that the remaining $H^i_{jk}$ must vanish provided that all the $Q_{ljk}$ do so. Hence, the problem of the vanishing of a tensor with infinitely many
components has been reduced to the vanishing of the coefficients of
a polynomial,
\begin{gather*}
Q_{ljk}=0, \\
\nonumber\\
\forall\quad l=0, \dots,4(N-2), \quad j,k=0,\dots,3N-5. \nonumber
\end{gather*}
\noindent
We look at these conditions
as a system on the derivatives of the Hamiltonian $h$. Using the antisymmetry of the Haantjes polynomial in $q$ and $r$, we can reduce the number of conditions,
since $Q$ is divisible by $(q-r)$. In the case $N=3$, the Haantjes polynomial reduces to
\begin{equation*}
Q(p,q,r)=\sum_{l=0}^{4}\sum_{m=0}^{4}\sum_{n=0}^{4}Q_{lmn}p^l
q^m r^n.
\end{equation*}
\noindent
We write
\begin{equation*}
Q(p,q,r)=(q-r)\sum_{l=0}^4M_l(q,r)p^l;
\end{equation*}
then, successively requiring the coefficients of $M_4$ and then $M_3$ to vanish
leads to $10$ partial differential equations of the form
\begin{equation*}
h_{ijk}=F_{ijk}\left(h_n,h_{nm},A^l\right), \qquad i,j,k,n,m,l=0,\dots,N-1.
\end{equation*}
\noindent
If these conditions hold, it is easy to verify directly that the Haantjes
polynomial $Q(p,q,r)$ is identically zero. We also recalculated the conditions
on the zeroth upper component
\begin{equation}\label{neccon}
H^0_{ij}=0
\end{equation}
\vskip 5mm
\noindent
which Ferapontov and Marshall used as \emph{necessary} conditions for the
Haantjes tensor to
vanish (see Section \ref{hych}). It is then straightforward to verify that
if these conditions (\ref{neccon}) hold, then $Q$ vanishes identically.
It thus follows that the necessary conditions are also sufficient, as Ferapontov
and Marshall had conjectured.

\section{Dubrovin-Novikov Hamiltonian formalism for hydrodynamic Vlasov equations}

Hydrodynamic Vlasov equations can be viewed as a generalization of systems
of hydrodynamic type. In particular, the Lie-Poisson bracket (\ref{lpb}) can be
seen as a Dubrovin-Novikov Poisson bracket (\ref{dnpb}). In this section we formalize an analogue of the DN Poisson bracket for these equations; and we provide two explicit examples, for a class of diagonal metrics and for
the second Hamiltonian structure of the Benney chain.  

The main objects for the construction of a bracket of this form has already
been defined: given a metric 
\begin{equation*}
g^{(p,q)}[f],
\end{equation*}
we can define the Christoffel symbols and the curvature
\begin{equation*}
b{p\choose q,r}, \qquad \quad  R{s \choose p, q, r},
\end{equation*}
given explicitly by (\ref{christ}), (\ref{curv}). Moreover, in view of the definition of the new
Poisson bracket, we notice that the elements $b^{ij}_k$ are here replaced
by
\begin{align*}
b{p,q \choose r}&=\int g^{(p,\alpha)}b{q \choose \alpha,r}d\alpha=\\
&=\frac{1}{2}\int\!\!\!\!\int g^{(p,\alpha)} g^{(q,\beta)}\left(\frac{\delta g_{(\alpha,\beta)}}{\delta f(r)}+\frac{\delta g_{(\beta,r)}}{\delta f(\alpha)}-\frac{\delta g_{(\alpha,r)}}{\delta f(\beta)}\right)d\alpha d\beta
\end{align*}

\vskip 10mm
\noindent
Given these objects, we can define an \emph{infinite dimensional Poisson bracket of hydrodynamic type} as

\begin{equation}\label{infpb}
\{K,H\}_g\!:=\!\!\int\!\!\!\!\int\!\!\!\!\int\!\!\!\!\int\!\!\!\!\frac{\delta K}{\delta f(p,x)}\left(g^{(p,q)}\frac{\partial}{\partial
x}+b{p,q \choose r}\frac{\partial f(r)}{\partial x}\right)\frac{\delta H}{\delta f(q,y)}dp dq dr dx,
\end{equation}
where $K,H$ are functionals of the type (\ref{hyfun}). Given a Poisson bracket of type (\ref{infpb}), the related Hamiltonian evolution equations are then
\begin{align*}
f_t(p)& =\{f,H\}_V=\\
& =\int g^{(p,q)}\frac{\partial}{\partial x}\frac{\delta H}{\delta f(q)}dq +\iint b{p,q \choose r}\frac{\partial f(r)}{\partial x}\frac{\delta H}{\delta f(q)}dq dr.\nonumber
\end{align*} 
\vskip 5mm
\noindent

\subsection{Diagonal metrics}
In order to find explicit expressions for the Christoffel symbols and of the
curvature, we now restrict ourselves to the case when the metric $g$
is diagonal with components $g_{(p,q)}$ of the form
\begin{equation}\label{diag}
g_{(p,q)}=\frac{1}{k[f]}\delta(p-q).
\end{equation}
The function $k[f]$ can depend on $f$ and finitely many of its derivatives with respect to $p$.
\noindent
The first advantage of a diagonal metric
is that the
inverse metric has components given by
\begin{equation}\label{invdiag}
g^{(p,q)}=k[f]\delta(p-q), 
\end{equation}
and so Hamilton's equations take the simpler form
\begin{equation*}
f_t(p,x)=k[f]\frac{\partial}{\partial x}\frac{\delta H}{\delta f(p)} +\iint b{p,q \choose r}\frac{\partial f(r)}{\partial x}\frac{\delta H}{\delta f(q)}dq dr.
\end{equation*} 

\noindent
For general $k[f]$, the calculation of the Christoffel symbols and of the
curvature presents many difficulties, due to the presence of higher $p$-derivatives
of $f$. On the other hand, the simplest case, when $k[f]$ depends only on
$f$ and not on its derivatives, turns out to be of relatively little interest, as
Hamilton's equations became the direct sum over $\mathbb{R}$ of one-dimensional
($N=1$) Poisson brackets of type (\ref{dnpb}). As an example, the $(\alpha-\beta)$ structure (\ref{alphabeta}), in the case
$\alpha=0$, belongs to this class, with
$$g^{(p,q)}=\frac{f(p)}{p}\delta(p-q).$$
\noindent
In this section, we develop
the first non-trivial case, when the function $k$ depends only on $f_p$.
We have the following
\begin{prop}
If for all $p,q$, the function $k$ depends only on on the first $p-$derivative
of $f$, i.e.
\begin{equation}\label{mfp}
g_{(p,q)}=\frac{1}{k[f_p]}\delta(p-q), 
\end{equation}
then the Christoffel symbols have the form
\begin{equation}\label{csdiag}
b{p \choose q,r}=\frac{k'[f_q]}{k[f_q]}\delta'(p-q)\delta(r-q)-\frac{1}{2}\frac{k''[f_q]}{k[f_q]}f_{qq}\delta(p-q)\delta(r-q),
\end{equation}
where $k'$, $k''$ are the first and second derivatives of $k$ respectively.
The metric $g$ is flat if and only if $k[f_p]$ is linear in $f_p$.
\end{prop}

\begin{proof}
The proof is a direct computation. In the calculation of the Christoffel symbols (\ref{christ}), we note that for the metric (\ref{mfp}) we have
\begin{equation*}
\frac{\delta g_{(p,q)}}{\delta f(r)}=\frac{k'[f_p]}{k[f_p]^2}\delta'(p-r)\delta(p-q),
\end{equation*}
substituting in the definition (\ref{christ}) of the Christoffel symbols,
this leads to
\begin{align*}
b{p \choose q,r}=&\frac{1}{2}\int k[f_p]\delta(p-\alpha)\left(\frac{k'[f_\alpha]}{k[f_\alpha]^2}\delta'(\alpha-r)\delta(q-\alpha)
+\frac{k'[f_\alpha]}{k[f_\alpha]^2}\delta'(\alpha-q)\delta(r-\alpha)+\right.\\
&-\left.\frac{k'[f_q]}{k[f_q]^2}\delta'(q-\alpha)\delta(r-q)\right)d\alpha=\\
=& \frac{1}{2}\frac{k'[f_p]}{k[f_p]}\delta'(p-r)\delta(p-q)+\frac{1}{2}\frac{k'[f_p]}{k[f_p]}\delta'(p-q)\delta(p-r)+\\
&-\frac{1}{2}\frac{k[f_p]k'[f_q]}{k[f_q]^2}\delta'(q-p)\delta(r-q).
\end{align*}
Rearranging, we obtain equation (\ref{csdiag}). For the calculation of the curvature (\ref{curv})
the technique is the same, and we obtain
\begin{align*}
R{s\choose p,q,r}&=\frac{k''[f_p]}{k[f_p]}\Big(\delta'(p-r)\delta'(p-s)\delta(p-q)-\delta'(p-q)\delta'(p-s)\delta(p-r)\Big)+\\
&+\frac{1}{2}\frac{k''[f_p]}{k[f_p]}\Big(\delta''(p-q)\delta(p-s)\delta(p-r)-\delta''(p-r)\delta(p-s)\delta(p-q)\Big)+\\
&+\frac{1}{2}\frac{k'''[f_p]}{k[f_p]}f_{pp}\Big(\delta'(p-q)\delta(p-s)\delta(p-r)-\delta'(p-r)\delta(p-s)\delta(p-q)\Big).
\end{align*}
\noindent
It is elementary to see that the condition $k''[f_p]=0$ leads to the vanishing
of the curvature tensor. On the other hand, evaluating the result above with suitable test functions, it is
possible to prove that the condition is also sufficient. 

\end{proof}

\vskip 7mm
\noindent
So, a metric of type (\ref{diag}) is flat if and only if has the form
\begin{equation*}
g_{(p,q)}=\frac{1}{a f_p+b}\delta(p-q),
\end{equation*}
with $a,b$ not depending on $f$. The related evolution equations are then

\begin{equation*}
f_t(p,x)=(a f_p+b)\frac{\partial}{\partial x}\frac{\delta H}{\delta f(p)} -a\frac{\partial f(p)}{\partial x}\frac{\partial}{\partial p}\frac{\delta H}{\delta f(p)},
\end{equation*} 
\noindent
In the special case $a=1$, $b=0$, we obtain the canonical Lie-Poisson bracket
(\ref{hve}), with Poisson operator of the form
\begin{equation}\label{1stbenney}
\pi^{(p,q)}=f_p\delta(p-q)\frac{\partial}{\partial x}+\int\delta'(p-q)\delta(p-r)f_x(r)dr.
\end{equation}

\vskip 5mm

\subsection{The second Hamiltonian structure for the Benney chain}
\noindent
The second Hamiltonian structure for the Benney equation (\ref{benney}) is
defined by the local Poisson operator
\begin{equation}\label{2ndbr}
(\Pi_2)^{kn}=(G_2)^{kn}\frac{\partial}{\partial x}+\sum_m(B_2)_m^{kn} A_x^m,
\end{equation}
where the metric $G_2$ has components,
\begin{align}\label{metg2}
(G_2)^{kn}= \,\,& k\,nA^{k-1}A^{n-1}+(k+n+2)A^{k+n}+\sum_{i=0}^{n-1}(k+i)A^{k+i-1}A^{n-i-1}+\nonumber\\
&-\sum_{i=0}^{n-2}(n-i-1)A^{k+i}A^{n-i-2},
\end{align}
and the Christoffel symbols are given by
\begin{align}\label{chrb2}
(B_2)^{kn}_m=&\, kn A^{k-1}\delta^{n-1}_m+(n+1)\delta^{k+n}_m-\sum_{i=0}^{n-2}\left(nA^{k+i}\delta^{n-i-2}_m\right)+\nonumber\\
&+\sum_{i=0}^{n-1}\left(iA^{n-i-1}
\delta^{k+i-1}_m+(k+i)A^{k+i-1}\delta^{n-i-1}_m\right).
\end{align}
\vskip 5mm
\noindent
The Hamiltonian density is $\frac{1}{2}A^1$.
This structure appeared for the first time in \cite{ku84}, where Kupershmidt
derived it as a dispersionless limit  of the second Poisson structure of the KP hierarchy. Recently, B{\l}aszak and Szablikowski rediscovered it (\cite{blsz02},\cite{sz06}) using the semiclassical R-matrix approach. Using the techniques developed in the previous sections, we obtain that the metric (\ref{metg2}) becomes,  in the Vlasov picture,
\vskip 1mm
\begin{align*}
g_2^{(p,q)}=& f_pf_q-pf_p\delta(p-q)+f(p)\delta(p-q)+\frac{f(p)f_q-f(q)f_p}{q-p}+\nonumber\\
& \delta(p-q)f_p\int\!\frac{f(r)}{r-p}dr-\delta(p-q)f(p)\int\!\frac{f(r)}{(r-p)^2}dr,
\end{align*}
and the Christoffel symbols (\ref{chrb2}) are then found to be
\vskip 5mm
\noindent
\begin{align*}
b_2{p,q\choose r}=& f_p\delta'(q-r)-q\delta'(q-r)\delta(p-r)+\frac{f(q)}{(q-p)^2}\delta(p-r)+\nonumber\\
& -\delta(p-q)\delta(p-s)\int\frac{f(s)}{(s-p)^2}ds+\nonumber\\ 
&+\delta'(q-p)\delta(p-r)\int\frac{f(s)}{s-p}ds+\\
&+\frac{f_p}{r-p}\left(\delta(p-q)-\delta(r-q)\right)+\frac{f(p)}{r-p}\left(\delta'(q-r)-\delta(q-p)\right).
\end{align*}
\vskip 5mm 
\noindent
In analogy with the Kupershmidt-Manin structure, if we consider a Hamiltonian
density depending on a finite number of moments $h=h(A^0,\dots,A^{N-1}),$
we obtain Vlasov equations of the form (\ref{chs}), with kernel $V{p\choose
q}$ given by
\begin{equation*}
\begin{split}
V{p\choose q}=&\sum_{n,m=0}^{N-1}h_{nm}q^m
\Bigg(-n A^{n-1}f_p-p^{n+1}f_p+p^nf(p)+\\
&\left.-\sum_{i=0}^{n-1}p^i A^{n-i-1}f_p+\sum_{i=0}^{n-2}(n-i-1)p^iA^{n-i-2}f(p)\right)+\\
&+\sum_{n=0}^{N-1}h_n
\left(-n f_p q^{n-1} +(n+1)p^n \delta(p-q)+\delta(p-q)\sum_{i=0}^{n-1}ip^iA^{n-i-1}+\right.\\
&\left.-f_p \sum_{i=0}^{n-1}p^i q^{n-i-1}-nf(p)\sum_{i=0}^{n-2}p^{i-1}q^{n-i-1}\right).
\end{split}
\end{equation*}
\vskip 5mm
\noindent
Consider now the Galilean transformation
\begin{equation*}
p\longmapsto p+\alpha,
\end{equation*}
where $\alpha$ is a constant. It is easy to verify that, under this change of coordinates, we have
\begin{equation*}
{\pi_2}^{(p+\alpha,q+\alpha)}={\pi_2}^{(p,q)}+\alpha{\pi_1}^{(p,q)},
\end{equation*}
where $\pi_1=\pi$ is the Poisson operator (\ref{1stbenney}). Thus the brackets (\ref{1stbenney})
and (\ref{2ndbr}) are compatible.

\section{Conclusions and open questions}
We have considered the problem of the integrability of Hydrodynamic chains which are Hamiltonian with respect to the Kupershmidt-Manin Poisson bracket. It turns
out that this problem can be reduced to the study of the corresponding hydrodynamic
Vlasov equation, for which the differential geometric objects related with
a chain become integral operators. Using this formulation, we calculated
the Haantjes tensor explicitly and found the conditions
for it to vanish, showing that the conditions found bt Ferapontov and Marshall
are in fact sufficient. In addition, we have constructed a suitable
Dubrovin-Novikov Hamiltonian formalism for hydrodynamic Vlasov equations,
getting explicit conditions for a class of diagonal metrics to be flat.
Finally, we have found the formulation, in Vlasov variables, of the second Hamitonian structure for the Benney hierarchy. It would be interesting to
study the analogous conditions on the Hamiltonian for the vanishing of the
Haantjes tensor for systems with this Hamiltonian structure.

\section*{Acknowledgments}
We are very grateful to E.V.Ferapontov and M.V.Pavlov for many fruitful discussions
about this problem. We would like to thank the European Commission's FP6
programme for support of this work through the ENIGMA network, and 
particularly its support of Andrea Raimondo.
Further support from the ESF through the MISGAM network is also gratefully
acknowledged.

\vskip 5mm
\bibliographystyle{unsrt}
\bibliography{GibRai07}

\def\cprime{$'$} \def\cprime{$'$} \def\cprime{$'$}
\begin{thebibliography}{10}

\bibitem{fema06}
E.V Ferapontov and D.G. Marshall.
\newblock {D}ifferential-geometric approach to the integrability of
  hydrodynamic chains: the {H}aantjes tensor.
\newblock arXiv:nlin.SI/0505013, 2005.

\bibitem{duno83}
B.A. Dubrovin and S.P. Novikov.
\newblock {H}amiltonian formalism of one-dimensional systems of the
  hydrodynamic type and the {B}ogolyubov-{W}hitham averaging method.
\newblock {\em Dokl. Akad. Nauk SSSR}, 270(4):781--785, 1983.

\bibitem{ts90}
S.P. Tsar{\"e}v.
\newblock The geometry of {H}amiltonian systems of hydrodynamic type. {T}he
  generalized hodograph method.
\newblock {\em Izv. Akad. Nauk SSSR Ser. Mat.}, 54(5):1048--1068, 1990.

\bibitem{pasvsh96}
M.V. Pavlov, S.I. Svinolupov, and R.A. Sharipov.
\newblock An invariant criterion for hydrodynamic integrability.
\newblock {\em Funktsional. Anal. i Prilozhen.}, 30(1):18--29, 1996.

\bibitem{ha55}
J.~Haantjes.
\newblock On {$X\sb m$}-forming sets of eigenvectors.
\newblock {\em Indagationes Mathematicae}, 17:158--162, 1955.

\bibitem{fekhmapa06}
E.V. Ferapontov, K.R. Khusnutdinova, D.G. Marshall, and M.V. Pavlov.
\newblock Classification of integrable {H}amiltonian hydrodynamic chains
  associated with {K}upershmidt$'$s brackets.
\newblock arXiv:nlin.SI/0607003, 2006.

\bibitem{be73}
D.J. Benney.
\newblock {S}ome properties of long nonlinear waves.
\newblock {\em Stud. Appl. Math.}, 52:45--50, 1973.

\bibitem{kuma77}
B.A. Kupershmidt and Ju.I. Manin.
\newblock Long wave equations with a free surface. {I}. {C}onservation laws and
  solutions.
\newblock {\em Funktsional. Anal. i Prilozhen.}, 11(3):31--42, 1977.

\bibitem{kuma78}
B.A. Kupershmidt and Ju.I. Manin.
\newblock Long wave equations with a free surface. {II}. {T}he {H}amiltonian
  structure and the higher equations.
\newblock {\em Funktsional. Anal. i Prilozhen.}, 12(1):25--37, 1978.

\bibitem{do93}
I.~Ya. Dorfman.
\newblock {\em Dirac structures and integrability of nonlinear evolution
  equations}.
\newblock Nonlinear Science: Theory and Applications. John Wiley \& Sons Ltd.,
  Chichester, 1993.

\bibitem{ku83}
B.A. Kupershmidt.
\newblock Deformations of integrable systems.
\newblock {\em Proc. Roy. Irish Acad. Sect. A}, 83(1):45--74, 1983.

\bibitem{pa06}
M.V. Pavlov.
\newblock {H}ydrodynamic chains and a classification of their {P}oisson
  brackets.
\newblock arXiv:nlin.SI/0603056, 2006.

\bibitem{pa206}
M.V. Pavlov.
\newblock Transformations of integrable hydrodynamic chains and their
  hydrodynamic reductions.
\newblock arXiv:nlin.SI/0604050, 2006.

\bibitem{pa306}
M.V. Pavlov.
\newblock Classification of integrable hydrodynamic chains and generating
  functions of conservation laws.
\newblock arXiv:nlin.SI/0603055, 2006.

\bibitem{pa406}
M.V. Pavlov.
\newblock The {H}amiltonian approach in classification and integrability of
  hydrodynamic chains.
\newblock arXiv:nlin.SI/0603057, 2006.

\bibitem{pa506}
M.V. Pavlov.
\newblock The {K}upershmidt hydrodynamic chains and lattices.
\newblock arXiv:nlin.SI/0604049, 2006.

\bibitem{gi81}
J.~Gibbons.
\newblock Collisionless {B}oltzmann equations and integrable moment equations.
\newblock {\em Phys. D}, 3(3):503--511, 1981.

\bibitem{ku84}
B.~A. Kupershmidt.
\newblock Normal and universal forms in integrable hydrodynamical systems.
\newblock In {\em Proceedings of the Berkeley-Ames conference on nonlinear
  problems in control and fluid dynamics (Berkeley, Calif., 1983)}, Lie Groups:
  Hist., Frontiers and Appl. Ser. B: Systems Inform. Control, II, pages
  357--378, 1984.

\bibitem{blsz02}
M.~B{\l}aszak and B.~M. Szablikowski.
\newblock Classical {$R$}-matrix theory of dispersionless systems. {II}.
  {$(2+1)$} dimension theory.
\newblock {\em J. Phys. A}, 35(48):10345--10364, 2002.

\bibitem{sz06}
B.~Szablikowski.
\newblock Private communication.

\end{thebibliography}

\end{document}